\documentclass[preprints,article,accept,moreauthors]{Definitions/mdpi} 

\firstpage{1} 
\pubvolume{xx}
\issuenum{1}
\articlenumber{5}
\pubyear{2020}
\copyrightyear{2020}
\history{Received: date; Accepted: date; Published: date}

\usepackage[T1]{fontenc}
\usepackage{graphicx}
\usepackage{amsmath}
{\left\lbrace\begin{array}{@{}l@{}}}%
{\end{array}\right.}

\Title{Cosmological perturbations in bouncing cosmologies and the case of the Pre-Big Bang scenario}

\Author{Valerio Bozza $^{1,2}*$}

\AuthorNames{Valerio Bozza}

\address{%
$^{1}$ \quad Dipartimento di Fisica “E.R. Caianiello”, Università di Salerno, Via Giovanni Paolo II 132,
I-84084 Fisciano, Italy\\
$^{2}$ \quad  Istituto Nazionale di Fisica Nucleare, Sezione di Napoli, Via Cintia, 80126, Napoli, Italy\\
}

\corres{Correspondence: valboz@sa.infn.it}

\abstract{The Pre-Big Bang cosmology inspired generations of cosmologists in attempts to cure the initial Big Bang singularity using a fundamental length scale as proposed by String Theory. The existence of a phase of collapse/inflation with increasing curvature followed by a cosmic bounce has been proposed as an alternative to standard inflation in the solution of the horizon and curvature problems. However the generation of a nearly scale-invariant spectrum of perturbations is not an automatic prediction of such scenarios. In this paper I review some general statements about the evolution of perturbations in bouncing cosmologies and some historically significant attempts to reconcile the predicted spectra with the observations. Bouncing cosmologies and in particular the Pre-Big Bang scenario stand as viable, although more complicated, alternatives to inflation that may still help solve current theoretical and observational tensions.
}


\begin{document}
\nolinenumbers
\section{Introduction}

Tracing the evolution of the Universe back to its origins is one of the hardest tasks in physics, involving General Relativity and Quantum Physics in regimes that cannot be reached by any conceivable experiments. At the same time, the Early Universe perhaps provides the only window offered by Nature to explore such regimes and look for possible observable consequences that may shed light on the foundations of the pillars on which our description of the world is based \cite{Agullo2012,Brandenberger2017,McAllister2008}. 
Hubble's law is our main observational evidences for the expansion of the Universe. If space-time participates in the dynamics of the matter it contains, it must be described through a theory that treats it as a dynamical entity. General Relativity provides a beautiful geometrical framework describing the evolution of space-time and its relation to the energy-momentum tensor. Taking the present expansion of the Universe as a matter of fact, the generic prediction of General Relativity for a Universe filled with ordinary matter and radiation is an initial singularity called Big Bang. Such simple model of the Universe suffers from some well-known problems \cite{Dodelson2003}:
\begin{itemize}
    \item The horizon problem: the Universe appears homogeneous on scales that only now are coming back in causal contact.
    \item The flatness problem: the Universe has little or no spatial curvature, requiring extremely special initial conditions.
\end{itemize}

A possible solution to these problems is a phase of accelerated expansion called inflation \cite{Guth1981,Linde1982,Albrecht1982}, which would stretch the space between particles to distances larger than the cosmological horizon, and also make the Universe spatially flat. At the end of the inflation, the Universe is essentially empty, save for quantum fluctuations that should seed the cosmological perturbations that later evolve into large-scale structures harboring galaxies and clusters of galaxies \cite{Bunch1978,Wang2014}.

In general, inflationary models typically assume that the accelerated expansion takes place at some energy scale below the Planck scale (e.g. at the Grand Unification \cite{Albrecht1982}), thus representing just a special phase within a Universe with ever-decreasing energy density and space-time curvature. In this case, any physics at energy scales higher than the inflation plays little or no role, as it is washed out by the following accelerated expansion. Therefore, any quantum gravity effects at the Planck scale and even the mere existence of the initial singularity remain hidden to our observations \cite{Vilenkin1983,Linde1984}.

It is widely believed that the Big Bang singularity is an accident of the extrapolation of General Relativity beyond its range of validity \cite{rovelli_2004}. When the curvature radius of the Universe becomes comparable with the Planck length, quantum effects should dominate the structure of space-time \cite{Amelino2013}. What happens at this scales depends on the unknown physics of quantum gravity. In this respect, String theory has played a major role in inspiring theoretical physicists about the possibilities opened by a quantum theory encompassing General Relativity \cite{Polchinski1998,Zwiebach2009}. Furthermore, the opportunity of finding any possible observational signatures of String theory through relics of an early stringy phase of the Universe has stimulated the research along many different directions \cite{McAllister2008}. One of the possibilities explored by string cosmologists in the last four decades is that the initial singularity might be cured by string theory and replaced by a finite maximum in the curvature and energy density. Then we can extend back the history of the Universe past this maximum in a new phase known as {\it Pre-Big Bang} \cite{Veneziano1991,Gasperini1993,Gasperini2003}. In this scenario, proposed by Gasperini and Veneziano in the nineties, the Universe started from an asymptotically flat space-time, where quantum fluctuations randomly grow and nucleate bubbles undergoing a super-inflation phase (an accelerated expansion with growing curvature) \cite{Buonanno1999}. When the curvature of these bubbles reaches the string scale (slightly below the Planck scale), string physics stops the super-inflation and gives rise to a decelerated expansion converging to the standard cosmological picture, including a radiation, a matter-dominated phase or any late time evolution.

Similarly to the Pre-Big Bang cosmology, many other string cosmologies were proposed thereafter in which the Big Bang is replaced by a cosmic bounce, i.e. a transition between a contraction era and the present expansion  \cite{BrandVafa1989,Khoury2001,Finelli2002} (for a review see \cite{Brandenberger2017}). Considering that the Pre-Big Bang superinflation becomes a contraction after a transformation of the action from the string frame to the Jordan frame, bouncing cosmologies can be viewed as a generalization of the Pre-Big Bang scenario, which was particularly motivated by the driving concept of T-duality \cite{Veneziano1991}.

The Pre-Big Bang and bouncing cosmologies solve the problems of standard cosmology similarly to standard inflation \cite{GaspVen2016}. However, important differences arise when we look at the cosmological perturbations. While standard inflation predicts a nearly scale-invariant spectrum both for scalar and tensor perturbations \cite{Dodelson2003}, the Pre-Big Bang cosmology (and other related string cosmologies) predicts steep blue spectra \cite{Brustein1995,Brustein21995}. As soon as the first observations of the Cosmic Microwave Background (CMB) in the first decade of the millennium revealed a nearly scale-invariant spectrum for scalar perturbations \cite{WMAP2003}, it became clear that a revision of the mechanism for the generation of cosmological perturbations was necessary in the Pre-Big Bang cosmology.

In this contribution in honour of the 70th birthday of Maurizio Gasperini, it is my pleasure to revive the studies of cosmological perturbations in bouncing cosmologies of the early years of the millennium in which I was personally involved. The legacy of those studies is still strong and continues to inspire theories and observations looking for stringy signatures in our sky. In Section 2 I present the Pre-Big Bang scenario and other bouncing cosmologies. In Section 3 I discuss the primordial spectra of perturbations generated in bouncing cosmologies. In Section 4 we face the delicate issue of how these perturbations evolve through the bounce. In Section 5 we will see how a scale-invariant spectrum can be re-generated after the bounce and check the observational and theoretical constraints. Finally we draw some conclusions in Section 6.

\section{The Pre-Big Bang phase}

The Pre-Big Bang cosmology assumes that at sufficiently low energies the space-time is fully described by the effective action of the bosonic sector of string theory \cite{Polchinski1998}:
\begin{equation}
S=-\frac{1}{2\lambda_S^{d-1}} \int \mathrm{d}^{d+1}x \sqrt{-g} e^{-\varphi} \left( R + \partial_\mu \varphi \partial^\mu \varphi +2 \lambda_S^{d-1}V(\varphi) -\frac{1}{12} H_{\mu \nu \alpha} H^{\mu \nu \alpha}  \right),  
\end{equation}
where $\lambda_S$ is the fundamental string length, $d$ is the number of spatial dimensions, $g_{\mu\nu}$ is the metric tensor, $R$ is the Ricci scalar, $\varphi$ is the scalar dilaton field, coming with a non-perturbative potential $V(\varphi)$, $H_{\mu \nu \alpha}= \partial_\mu B_{\nu \alpha}+ \partial_\nu B_{\alpha \mu} + \partial_\alpha B_{\mu \nu}$ is the field strength of the Kalb-Ramond antisymmetric field $B_{\mu\nu}$.

String theory is consistently formulated with $d=9$ or $d=10$ (in the M-theory version). We can thus imagine that the Pre-Big Bang phase leads to an expansion of only three spatial dimensions accompanied by a contraction of the remaining 6 dimensions (if we adopt $d=9$), whose volume shrinks down to the string scale. Therefore, we split the metric in the following form
\begin{equation}
    ds^2=a^2(\eta)d\eta^2-a^2(\eta)\delta_{ij}dx^idx^j - b^2(\eta) \delta_{lm}dy^l dy^m, \label{splitmetric}
\end{equation}
where $\eta$ is the conformal time, $x^i$ ($i=1,2,3$) are the coordinates along the three large dimensions, $y^l$ ($l=4, \ldots, 9$) are the coordinates in the internal small dimensions. $a(\eta)$ is the scale factor for the large dimensions and $b(\eta)\equiv e^{\beta(\eta)}$ is the scale factor describing the contraction of the small dimensions. We have assumed homogeneity and isotropicity within each of the two blocks of coordinates. In this framework, we may introduce an effective 4-dimensional dilaton $\phi=\varphi-6\log b$ and write the antisymmetric field strength in terms of a single pseudoscalar axion field $\sigma$ as $H^{abc}=e^{\phi}\epsilon^{abcd} \partial_d \sigma$ \cite{Gasperini2003}.

The dilaton is non-minimally coupled to the metric in this action. However, by a conformal transformation of the metric, we can go from the physical string frame to an equivalent Einstein frame in which the dilaton is minimally coupled. The new metric is related to the old one by
\begin{equation}
    \tilde{g}_{\mu\nu} =g_{\mu\nu}e^{-\phi}.
\end{equation}

In the Einstein frame, the description of the Pre-Big Bang phase is particularly simple, since it becomes an accelerated contraction driven by the kinetic energy of the dilaton field. The exact dynamics, however, depends on the evolution of the spatial dimensions. The equations for the metric, dilaton and axion are
\begin{eqnarray}
 {\cal H}^2 &=& \frac{1}{12} \left(\phi'^2 + 12 \beta'^2 + e^{2\phi} \sigma'^2 \right) \\
 {\cal H}' +2{\cal H}^2 & = & 0 \\
 \beta''+2{\cal H} \beta' & = & 0 \\
 \phi'' +2{\cal H} \phi' & = & e^{2\phi \sigma'^2} -\frac{dV}{d\phi}\\
 \sigma'' + 2 {\cal H} \sigma ' & = & -2\phi' \sigma', \label{Eqaxionbkg}
\end{eqnarray}
where ${\cal H}\equiv \tilde a'/ \tilde a$ and the prime denotes derivative with respect to the conformal time.

If the axion does not contribute to the cosmic background in the Pre-Big Bang phase and the dilaton has a vanishing potential, then we have the following solutions
\begin{eqnarray}
\tilde a &\sim & |\eta|^{1/2}  \label{atilde}\\
\beta & \sim & |\eta|^{s}  \label{beta}\\
e^\phi & \sim & |\eta|^{\pm \sqrt{3}\sqrt{1-4s^2}}. \\
\sigma & \sim & \sigma_*
\end{eqnarray}

Since the dilaton and the moduli field $\beta$ enter as degenerate scalar fields, the contraction rate of the internal dimensions $s$ remains as a free parameter. The solution for the dilaton depends on this parameter and may be growing or decaying depending on the chosen sign in the exponent. 

These solutions are singular for $\eta=0$ and may describe either a Pre-Big Bang ($\eta<0$) or a Post-Big Bang ($\eta>0$). It is believed that the full string theory removes the singularity and ensures that the Pre-Big Bang contraction bounces to a standard Post-Big Bang evolution. It is also assumed that the same string phase is characterized by a non-trivial potential for the dilaton, which is frozen in the Post-Big Bang so that the effective gravitational coupling $M_S e^{-\phi/2}$ remains fixed to the observed Planck mass $M_P$.

The Pre-Big Bang phase, then, appears as an accelerated contraction governed by Eq. (\ref{atilde}) in the Einstein frame. From the point of view of the string frame, instead, we have
\begin{equation}
    a= \tilde a e^{\phi/2} \sim |\eta|^{\frac{1}{2}(1\pm \sqrt{3}\sqrt{1-4s^2})},
\end{equation}
which corresponds to an accelerated expansion (pole inflation) in the negative sign branch if $s<1/\sqrt{6}$. Such superinflation is able to solve the problems of standard cosmology and may represent a good alternative to slow-roll inflation, if we only consider the background evolution. In the Einstein frame, the exponent of the scale factor is fixed to $1/2$, whatever the relative contraction rate of the internal dimensions. Alternative bouncing cosmologies have been proposed with different contraction rates, depending on the field content and the geometry of the space-time \cite{Khoury2001,Finelli2002}. In the next section, we will leave this exponent as a free parameter so as to derive the primordial spectra of cosmological perturbations for a general class of bouncing cosmologies. We will therefore set
\begin{eqnarray}
 && \tilde a \sim |\eta|^{q_-} \label{generic atilde}\\
 && \mathcal H \sim \frac{q_-}{\eta}
\end{eqnarray}
in the pre-bounce phase and go back to $q_-=1/2$ for the Pre-Big Bang. We note that the space-time curvature vanishes at $\eta \rightarrow -\infty$. Therefore, all these models start from an asymptotically flat past \cite{Buonanno1999}.

It has been noticed that fast contractions with $q_->1/2$ are exposed to an uncontrolled growth of anisotropies that may lead to chaotic mixmaster oscillations \cite{Damour2000}. The Pre-Big Bang scenario is just at the divide between safe and unsafe backgrounds. A possible solution to this problem was proposed in Ref. \cite{Bruni}. 

\section{Primordial spectra in bouncing cosmologies}

In order to justify the existence of inhomogeneities in the present Universe in the form of large-scale structures, clusters of galaxies and ultimately to all gravitationally bound structures we see today, we must follow the evolution of cosmological perturbations from the initial seeds to the present time. In general, perturbations to the Robertson-Walker metric are divided in three classes according to their behavior under rotations of the spatial rotations \cite{Mukhanov1992}. Vector perturbations grow in the pre-bounce phase but decay and become negligible in the post-bounce expansion \cite{Giovannini2004}. We will thus focus on tensor and scalar perturbations.

\subsection{Tensor perturbations}

Tensor perturbations are gauge invariant under generic coordinate transformations. They carry two degrees of freedom corresponding to the two polarization states of gravitational waves. They follow the wave equation
\begin{equation}
    h''_{ij}+2{\cal H} h'_{ij} + k^2 h_{ij} = 0,
\end{equation}
where we have gone to Fourier space, so that $\nabla^2 \rightarrow -k^2$, and $k$ is the wave number of the Fourier mode.

Since bouncing cosmologies start from an asymptotically flat spacetime, it is assumed that such flat space is only populated by vacuum fluctuations that eventually grow as the inflation/contraction begins. The pre-bounce solution is thus normalized to the initial vacuum state defined in terms of the canonically normalized field 
\begin{equation}
    \tilde h_{ij}= \frac{1}{\sqrt{2} \kappa} \tilde a h_{ij},
\end{equation}
with $\kappa=8\pi G$. The solution will depend on the specific background chosen through the scale factor $\tilde a$ as parameterized through Eq. (\ref{generic atilde}). In definitive, we have
\begin{equation}
h_{ij}=\frac{c_h}{\tilde a} \sqrt{|\eta|} H^{(1)}_\nu (k \eta), \label{tensor}
\end{equation}
where $c_h$ is a normalization constant, $H^{(1)}_\nu$ is the Hankel function of the first kind and $\nu=\frac{1}{2}-q_-$.

As the pre-bounce contraction proceeds, more and more Fourier modes will exit the Hubble horizon as $|k\eta|$ becomes less than one. In the super-horizon limit $|k\eta| \ll 1$, Eq. (\ref{tensor}) gives
\begin{equation}
h_{ij}\sim c_{h1} k^\nu |\eta|^{1-2q_-} +c_{h2} k^{-\nu}. \label{tensor SH}
\end{equation}
Therefore, two modes exist. The first mode becomes negligible for $q_-<1/2$ (slow contraction) leaving a spectrum dominated by the constant mode. For the Pre-Big Bang scenario ($q_-=1/2$), the first mode is replaced by a logarithmic growth, and both modes have power $\nu=0$ \cite{Brustein1995}.

The power spectrum of tensor fluctuations is defined by
\begin{equation}
P_h(k) \sim k^3 |h_{ij}|^2 \sim k^{n_T}    
\end{equation} 
in terms of a spectral index $n_T$.

With the super-horizon limit (\ref{tensor SH}), we have $n_T=3-2\nu=2+2q_-$. Any contracting Universe will thus have a steep blue spectrum for tensor modes \cite{Brustein1995,Khoury2001}. The Pre-Big Bang scenario, in particular, has $n_T=3$. A slow-roll inflation, instead, would be characterized by $\tilde a=-1/\eta$, corresponding to $q_-=-1$, which gives $n_T=0$, a scale-invariant spectrum.

\subsection{Scalar perturbations}

Scalar perturbations are defined by the metric
\begin{equation}
    ds^2=\tilde a^2\left\{(1+2A) d\eta^2-2B_{,i} d\eta dx^i - \left[(1-2\psi)\delta_{ij}+ 2E_{,ij}\right]dx^i dx^j \right\}.
\end{equation}

A generic perturbed energy-momentum tensor compatible with this metric is
\begin{equation}
    T_{\mu}^\nu=\left( \begin{tabular}{cc}
         $\rho + \delta \rho$ &  $-(\rho+p)u_{,i}$\\
         $(\rho+p)u_{,i}$ & $-(p+\delta p) \delta_{ij} - \xi_{,ij}$ 
    \end{tabular}\right),
\end{equation}
where $\rho$ and $p$ represent the total energy density and pressure with their perturbations $\delta \rho$ and $\delta p$, $\xi$ is the anisotropic stress and $u$ is the scalar velocity potential for the cosmic fluid. This energy-momentum tensor can be specialized to the cosmological model of interest. In principle, any corrections to the Einstein equations arising in the specific theory of gravity used to describe space-time can be incorporated in the energy-momentum tensor on the right hand side of the Einstein equations. In this way, we can use the perturbed Einstein equations to follow scalar perturbations and discuss the possible outcomes depending on the effective energy-momentum content. With this spirit, we can introduce the following gauge invariant combinations
\begin{eqnarray}
&& \Psi\equiv\psi +\mathcal{H}(E'-B) \\
&& \zeta\equiv\psi+ \mathcal{H}u \\
&& \delta \rho_u \equiv \delta \rho - \rho' u\\
&& \delta p_u \equiv \delta p - p' u
\end{eqnarray}
and write the following independent equations after eliminating $A$ from the component $(0i)$ of the Einstein equations \cite{Bozza2006}:
\begin{eqnarray}
 2 \nabla^2 \Psi& = & \tilde a^2 \delta \rho_u \label{Poisson} \\
 \frac{2 \left( \mathcal{H}^2 -\mathcal{H}' \right)}{ \mathcal{H}} \zeta' & = & \tilde a^2\nabla^2 \xi-  \tilde a^2\delta p_u \label{Eqzeta'}\\
 \Psi'+\frac{ 2\mathcal{H}^2 -\mathcal{H}' }{ \mathcal{H}} \Psi-\frac{ \mathcal{H}^2 -\mathcal{H}' }{ \mathcal{H}} \zeta  & = & a^2 \mathcal{H} \xi, \label{EqPsi'}
\end{eqnarray}

Each of these gauge invariant variables has a specific physical meaning that becomes apparent in a comoving gauge, where $\delta \rho_u$, $\delta p_u$ and $\zeta$ can be identified with the energy density, pressure and spatial curvature on comoving hypersurfaces respectively. The behavior of scalar perturbations depends on the specific matter content dominating the inhomogeneities on the right hand side. In general, the physics describing such matter will be expressed by equations of state relating $\delta p_u$ and $\xi$ to $\delta \rho_u$. For an asymptotically flat background, we can assume that the perturbations behave in some simple way. Therefore, we set $\delta p_u=c_-^2 \delta \rho_u$ and $\xi=0$. These relations hold e.g. for a perfect fluid or for a scalar field and are sufficiently generic to encompass all relevant cases. Then Eqs. (\ref{Poisson})-(\ref{EqPsi'}) can be combined into a second order equation for $\zeta$, which is related to the Sasaki-Mukhanov variable that puts the perturbed action in a canonical form. The solution is 
\begin{equation}
    \zeta=C_\zeta \frac{\mathcal{H}}{\tilde a \sqrt{\mathcal{H}^2 -\mathcal{H}'}} \sqrt{|\eta|} H_\nu^{(1)}(c_- k |\eta|),
\end{equation}
where $C_\zeta$ is a normalization constant and the other notations are the same as for Eq. (\ref{tensor}). This solution can be expanded for small arguments $k|\eta|\ll 1$ to find the behavior of modes outside the Hubble horizon at the bounce. The other scalar potentials can be obtained from $\zeta$ through Eqs. (\ref{Poisson})-(\ref{EqPsi'}). In particular the Bardeen potential is obtained by Eq. (\ref{Eqzeta'}), which approximates to $\Psi \sim \zeta'/(k^2|\eta|)$. We then have \cite{Bozza2006}
\begin{eqnarray}
 && \zeta \sim c_{\zeta1} k^\nu |\eta|^{1-2q_-} + c_{\zeta2}k^{-\nu} \\
 && \Psi \sim c_{\Psi1}k^{\nu-2} |\eta|^{-1-2q_-} + c_{\Psi2}k^{-\nu}.
\end{eqnarray}
Note that the constant mode in $\zeta$ is killed by the derivative appearing in Eq. (\ref{Eqzeta'}). The higher order term in the expansion of the Hankel function in $\zeta$ is  $k^{-\nu+2}|\eta|^2$, which generates a constant mode in $\Psi$ similar to the one in $\zeta$. 

These asymptotic expansions describe the behavior of scalar perturbations outside the horizon in the approach to the cosmic bounce. It is interesting to note that the constant modes have the same spectrum as the tensor modes, which is in general steeply blue, as discussed in the previous subsection \cite{Brustein21995}. The other mode is decaying for $\zeta$  but is fast growing for $\Psi$. This growing mode would give rise to a spectrum
\begin{equation}
P_\Psi(k) \sim k^3 |\Psi|^2 \sim k^{n_s-1}    
\end{equation} 
with a spectral index $n_s=4+2\nu-4=1-2q_-$. For the Pre-Big Bang scenario ($q_-=1/2$) this would be a red spectrum, but it is interesting to note that in the limit of slow contraction $q_-\ll 1$ this mode becomes nearly scale-invariant. For this reason, very great attention was given to models proposing a slow contraction before the bounce. One of these was certainly the Ekpyrotic/cyclic model \cite{Khoury2001,Khoury2002,Lehners2007,Buchbinder2007,Ijjas2014,Steinhardt2002}, in which the Big Bang is interpreted as the collision of our visible Universe, represented by a 3-brane with a hidden brane travelling across the extra-dimensions. Then, the pre-bounce phase can be described by a slow contraction in which the scalar perturbations develop a scale-invariant spectrum via the growing mode of $\Psi$. However, the optimistic view in which this spectrum determines the fluctuations at horizon re-entry in the post-bounce has been questioned in several works and has raised a considerable debate \cite{Brandenberger2001,Lyth2002,Martin2002,Tsujikawa2002,Hwang2002,Durrer2002,Cartier2003}. We will come back to this issue in the next section. Finally, we note that a scale-invariant spectrum in $\zeta$ can be obtained for a dust-dominated contraction $q_-=2$, which turns the decaying mode into a growing mode with a spectral index $n_s=4+2\nu=5-2q_-$ \cite{Finelli2002}. However, it is not clear how to avoid that sooner or later other sources (e.g. radiation or scalar fields) dominate over dust as the contraction leads toward higher and higher densities \cite{Damour2000,Karouby2010}. 

\subsection{Axion perturbations}

The study of the Pre-Big Bang framework is not concluded by the dominant fields: the metric tensor and the dilaton. We have seen that the antisymmetric 2-form can be recast in terms of a pseudoscalar axion $\sigma$, which does not take part in the cosmic background evolution because its contribution is negligible compared to the dilaton \cite{Copeland1997}. The axion perturbations (here indicated by the symbol $\chi$) will then follow an equation similar to Eq. (\ref{Eqaxionbkg}), with the complement of the spatial derivatives
\begin{equation}
 \chi'' + 2 ({\cal H}+\phi') \chi ' +k^2 \chi = 0.
\end{equation}

The solution of this equation with the normalization to vacuum fluctuations in the asymptotic past is
\begin{equation}
\chi=\frac{c_\sigma}{\tilde a e^\phi} \sqrt{|\eta|} H^{(1)}_r (k \eta), \label{axion}
\end{equation}
where $c_\sigma$ is a normalization constant and $r=\sqrt{3}\sqrt{1-4s^2}$.

The asymptotic expansion for small arguments, valid in the approach to the bounce as modes exit the horizon, is
\begin{equation}
\chi \sim c_{\sigma1} k^{-r} |\eta|^{-2r} +c_{\sigma2} k^{r}. \label{axion SH}
\end{equation}
The first mode dominates at the bounce and determines the spectrum of the axion as
\begin{equation}
P_\sigma(k) \sim k^3 |\chi|^2 \sim \frac{M_S^2}{M_P^2}k^{n_\sigma-1}   \label{axion spectrum}  
\end{equation} 
in terms of a spectral index $n_\sigma=4-2r=4-2\sqrt{3}\sqrt{1-4s^2}$. Therefore, the type of spectrum for the axion fluctuations generated in the Pre-Big Bang phase is related to $s$, i.e. the rate of compactification of the internal dimensions (see Eq. \ref{beta}) \cite{Copeland1997,Melchiorri1999}. The minimal value for the spectral index is obtained for stable internal dimensions $s=0$, corresponding to $n_\sigma=4-2\sqrt{3}=0.54$. The maximum value is obtained when $s=1/2$, i.e. when the rate of contraction of the internal dimensions equals that of the large three dimensions. In this limit, the dilaton becomes constant and the axion spectrum has the same spectral index as the other scalar perturbations $n_\sigma \rightarrow 4$. In-between these two extrema, a possibility for a scale-invariant spectrum arises if $s=1/4$. This particular situation corresponds to the case in which the contraction rate of the volume of the 3 external dimensions perfectly matches the contraction rate of the volume of the 6 internal dimensions in the Einstein frame:
\begin{equation}
R= \frac{\tilde a ^3}{e^{6\beta}}=\frac{|\eta|^{3/2}}{|\eta|^{6s}} \rightarrow 1 \;\; \mathrm{for} \; \; s \rightarrow \frac{1}{4}.
\end{equation}

Of course, the freedom left by the possible backgrounds is limited by theoretical and observational constraints that will be discussed after the whole evolution of the perturbations across the bounce will be presented in the next section.

We note that also in the Ekpyrotic scenario it is possible to add a second scalar field acquiring a scale-invariant spectrum in the pre-bounce phase, as suggested in the New Ekpyrotic scenario \cite{Notari2002,Lehners2007}. 

\section{Across the bounce}

The evolution of tensor perturbations in bouncing cosmologies is quite straightforward, since nothing is expected to happen to modes outside the horizon at the bounce. The spectrum generated at the onset of the bounce should pass unaffected and thus a steep blue spectrum is expected in all models, including the Pre-Big Bang scenario \cite{Brustein1995}. This prediction marks a sharp distinction with respect to standard inflationary models, where the tensor spectrum is always nearly scale-invariant or just slightly red. So, if any contribution of tensor modes to low multipoles of the CMB is ever measured, bouncing cosmologies would be ruled out or need to be completely reformulated. On the other hand, the blue spectrum opens the possibility to measure a stochastic gravitational wave background at high frequencies, accessible to future interferometers \cite{Gasperini2016}.

Scalar perturbations are more tricky, since they directly depend on the energy-momentum source dominating the Universe. In order to step out of the initial collapse and avoid a big crunch singularity, some new physics is invoked to intervene at the quantum gravity scale. In the Pre-Big Bang scenario, for example, the string mass plays the role of a cut-off for the curvature and the energy density \cite{Gasperini2003}. The string effective action should be complemented by higher order terms that should drive the Universe toward an expansion phase. As an interesting alternative, an explicit realization of the bounce through a non-local potential for the dilaton has been presented in Refs. \cite{GGV2003,GGV2004}. 

Non-singular bouncing models have been obtained in many ways by adding a cosmological component violating the null energy condition at the bounce \cite{Finelli2003,Allen2004,BV12005,BV22005,Battefeld2006,Biswas2006,Finelli2008}. In the Ekpyrotic/cyclic model, the collision of our brane with the hidden brane is believed to occur through a singularity of the metric, in the sense that the distance between the two branes vanishes at the bounce \cite{Khoury2001}.

In any case, one may describe the evolution of the Universe during the bounce by complementing the Einstein equations by additional terms arising in this high-energy regime. These terms can be moved to the right hand side and act as an effective additional source. In this respect, the violation of the Null Energy Condition (NEC), which is needed to convert a collapse phase to an expansion, can be performed by these additional effective terms. Therefore, the generation of possible ghosts and instabilities due to negative energy density can be avoided if the NEC violation is due to these higher order terms. Explicit realizations of ghost-free bouncing scenarios have been indeed presented \cite{GGV2003,GGV2004,Biswas2006}.

We may guess that the evolution of scalar perturbations across the bounce may be heavily affected by these additional terms, which may appear as effective sources in the perturbation equations (\ref{Poisson})-(\ref{EqPsi'}). However, this is not necessarily the case. In fact, inhomogeneities in the sources dominating the bounce necessarily descend from the inhomogeneities generated all along the previous pre-bounce collapse. Leaving apart isocurvature perturbations generated by subdominant fields in the pre-bounce, if we wish to study the evolution of the adiabatic mode, we assume that the source terms in Eqs. (\ref{Eqzeta'})-(\ref{EqPsi'}) are functions of $\zeta$ and $\Psi$ with the addition of possible powers of $k^2$ from Laplacian operators:
\begin{eqnarray}
 && \delta p_u = F(k^2) \zeta + G(k^2) \Psi  \label{p-bounce}\\
 && \xi =J(k^2) \zeta + K(k^2) \Psi,
\end{eqnarray}
with $F$, $G$, $J$ and $K$ being regular power expansions in $k^2$ \cite{Bozza2006}.

At this point, the evolution of the gauge-invariant variables $\zeta$ and $\Psi$ is fully contained in the closed set of equations (\ref{Eqzeta'})-(\ref{EqPsi'}). Although we do not know the details of the background evolution during the bounce, we may identify the modes surviving after the bounce and their time dependence by putting the equations in the integral form and solving them recursively, as proposed in Ref. \cite{Bozza2006}. The only two scales in the problem are the wave number $k$ and the fundamental bounce scale $H_i$, which can be identified as the scale at which the new physics triggering the bounce comes into play. This scale governs the bounce duration, energy density and curvature. So, each integral over the conformal time $\eta$ in the formal solutions of Eqs. (\ref{Eqzeta'})-(\ref{EqPsi'}) will introduce a new factor of $H_i$ in the expressions without changing the $k$ dependence.

With these rules, it is then possible to write down the post-bounce behavior of the scalar perturbations as \cite{Bozza2006}

\begin{eqnarray}
 && \zeta \sim d_{\zeta1} k^\nu |\eta|^{1-2q_+} + d_{\zeta2}k^{-\nu} + d_{\zeta3}k^{\nu}+ d_{\zeta4}G_0 k^{\nu-2} \\
 && \Psi \sim d_{\Psi1}k^{\nu-2} |\eta|^{-1-2q_+} + d_{\Psi2}k^{-\nu} + d_{\Psi3}k^{\nu}+ d_{\Psi4}G_0k^{\nu-2},
\end{eqnarray}
where the $d_{XX}$ are constants possibly related to the fundamental bounce scale $H_i$. We have also parameterized the post-bounce scale-factor as $\tilde a \sim \eta^{q_+}$, in analogy with Eq. (\ref{generic atilde}). The first two modes coincide with the dominant modes generated in the pre-bounce. However, since the contraction has been converted to an expansion, the growing mode of $\Psi$ is now a decaying mode \cite{Lyth2002}. The final spectrum will be thus dominated by the constant modes. Leaving apart the last term for the moment, the dominant contribution comes from the third mode, which gives rise to the spectral index $n_s=4+2\nu$ already discussed in the pre-bounce context as yielding a scale-invariant spectrum in the limit of a dust-like contraction \cite{Finelli2002}. The last terms contains a factor $G_0$ remarking the fact that it only exists if the $G(k^2)$ function introduced in Eq. (\ref{p-bounce}) is non-zero for $k^2 \rightarrow 0$ \cite{Bozza2006}. Therefore, only if during the bounce there is a dependence $\delta p_u \sim \Psi$ rather than $\delta p_u \sim k^2\Psi$, the spectrum of the growing mode of $\Psi$ is transferred to a constant mode and survives after the bounce. 

This condition was already stated in a different way by modeling the bounce as a thin space-like hypersurface and applying Israel junction conditions \cite{Durrer2002}. It clarifies a long-debated issue about the viability of Ekpyrotic/cyclic models as an alternative to standard inflation for the generation of the primordial spectrum of scalar perturbations. A slow contraction indeed generates a scale-invariant spectrum which matches to a decaying mode in the post-bounce, unless we have some unconventional source proportional to $\Psi$ rather than $k^2 \Psi$ in the spatial Einstein equations. No explicit example of such sources has been provided up to now. Regular bounces obtained by perfect fluids or scalar fields only involve sources with $\delta p_u \sim \delta \rho_u$, which is bound to be proportional to $k^2 \Psi$ by Eq. (\ref{Poisson}). Therefore, these toy models have confirmed that the original growing mode of $\Psi$ decays in the post-bounce expansion \cite{BV12005,BV22005,Allen2004,Finelli2008}. However, it might still be possible that some new physics mechanisms may replace the $k^2$ factor by another scale in the problem, being the bounce scale $H_i$ itself or some geometric scale, such as the size of the extra-dimensions or similar. Therefore, it would still be interesting to continue the search for fully self-consistent bouncing cosmologies in which the same fields dominate the pre-bounce background and generate the observed spectra. Otherwise, we have no other route than looking at the perturbations of sub-dominant fields.

\section{Re-generation of cosmological perturbations by the axion/curvaton}

Any credible alternative to standard inflation should contain a mechanism to generate fluctuations compatible with CMB and large-scale structure observations. This requires a nearly scale-invariant adiabatic primordial spectrum of scalar perturbations with the correct amplitude.

As a result of the studies outlined in the previous sections, we have seen that the Pre-Big Bang cosmology and other related bouncing cosmologies inspired by string theory or other quantum gravity theories are not able to naturally generate a nearly scale-invariant spectrum for scalar perturbations if we confine our attention to the dominant fields. However, we have also noticed that the axion field of the Pre-Big Bang scenario, although subdominant, typically develops a perturbation spectrum with a slope spanning a relatively wide range of possibilities depending on the specific background evolution \cite{Copeland1997}. In particular, a scale-invariant spectrum is obtained for a particularly symmetric background in which the contraction rates of internal and external dimensions is the same.  However, even in the case that such fluctuations dominate in some regime, they would only give rise to isocurvature fluctuations, which are severely constrained by current observations.

A possible mechanism to convert isocurvature perturbations to adiabatic ones has been suggested through the so-called curvaton field \cite{LythWands2002}. The idea is that the field responsible for the standard inflation (or the collapse/inflation for bouncing cosmologies) is not the same that generates the observed fluctuations. A second scalar field takes over as the dominant field at some point in the cosmic history and then imprints its fluctuations on $\zeta$ and $\Psi$. The resulting perturbations would be adiabatic as they come from the new dominant source. At the end of its lifetime, the curvaton decays to radiation fields and the standard expansion history proceeds without other changes. 

As soon as the curvaton mechanism was proposed, it was clear that the Pre-Big Bang axion would make an ideal candidate to implement the curvaton in a practical case \cite{Enqvist2002}. A scalar field driven by its kinetic energy redshifts faster than radiation and would never dominate in an expanding Universe. For a scalar field with a potential, we may consider two different regimes: if the potential dominates and the kinetic energy is negligible, the scalar field behaves as a cosmological constant and gives rise to an exponential expansion; if the scalar field oscillates at the bottom of its potential, the average expansion rate tends to be similar to a matter-dominated phase with $a \sim \eta^2$.

In the Pre-Big Bang scenario, it is believed that the dilaton field is stabilized by a non-perturbative potential acquired during the string phase. Similarly, the axion may receive a non-perturbative potential and find itself displaced from the minimum. So, as the expansion of the Universe starts with an early radiation phase, the axion slowly rolls down its potential. At some point, it will dominate over radiation. Depending on its initial value, this may happen when it is still in a slow-roll phase or when it has already reached the minimum of its potential and started its oscillations. It is in this phase that the axion fluctuations dominate and become adiabatic. As the axion finally decays, its fluctuations are naturally inherited by the new radiation field, which will dominate the following era.

Although the route for the curvaton/axion mechanism is clear, many details need to be worked out explicitly to check that no physical constraints are violated \cite{BGGV12002,BGGV22003}.

Let us assume that the initial value of the axion at the onset of the Post-Big Bang expansion is $\sigma_i$. At this time, the dilaton is already frozen at the present value and the extra-dimensions are finally stabilized, so that the space-time dynamics is effectively governed by the Friedmann equations. Moreover, with a fixed dilaton, there is no distinction between Einstein and string frames, so $a \sim \tilde a$ up to a constant. We will thus use the symbol $a$ for the scale factor without the tilde in the Post-Big Bang. The energy-momentum tensor is initially dominated by a radiation field, while the axion is still subdominant. However, a non-perturbative potential $V(\sigma)$ arises for the axion, so that the background evolution is described by the following set
\begin{eqnarray}
 {\cal H}^2 &=& \frac{a^2}{6} \left(\rho_r + \rho_\sigma \right) \\
 {\cal H}'  & = & - \frac{a^2}{12} \left(2\rho_r + \rho_\sigma+3 p_\sigma \right) \\
 \sigma'' + 2 {\cal H} \sigma ' & = & - \frac{dV}{d\sigma}, \label{Eqaxionbkg2}
\end{eqnarray}
with $\rho_\sigma = \frac{\sigma'^2}{2a^2} +V$ and $p_\sigma = \frac{\sigma'^2}{2a^2} -V$.

The potential for the Kalb-Ramond axion arising after the string phase is believed to be periodic. However, close to a minimum, it makes sense to deal with a quadratic approximation $V\simeq \frac{1}{2}m^2 \sigma^2$. Then the axion starts from some initial value $\sigma_i$ such that $\rho_r \gg \rho_\sigma$. If the axion kinetic energy is initially negligible, we have a slow roll phase for the axion. As the axion reaches the minimum of its potential, it starts to oscillate. This occurs when the expansion rate in cosmic time $H  \equiv a \mathcal{H} =  H_{osc}\sim m$. During oscillations, the energy density of the axion redshifts as $a^{-3}$, similarly to pressureless matter. 

Either as a slowly rolling scalar field or as pressureless matter, the axion will sooner or later dominate over the early radiation, which redshifts as $a^{-4}$. The axion dominance will occur at $H=H_{\sigma}=m \sigma(t)$. Depending on the initial value of the axion $\sigma_i$, the axion phase may start either before or after the first oscillation. In particular, we have an early dominance of the axion if $\sigma_i>1$ (in Planck units) and a late dominance otherwise. The two cases have different background evolutions and must be treated separately.

Finally, the axion will decay through its gravitational coupling to photons when the space-time curvature becomes of the order of the decay rate at $H=H_d \sim \frac{m^3}{M_P^2}$. After this phase, the standard radiation era takes place as usual and will inherit the fluctuation spectrum of the axion as its adiabatic mode. The successful transfer of the initial spectrum of the axion to the re-generated radiation has been tested analytically and numerically in Ref \cite{BGGV22003}.

The space allowed for an axion dominated phase in the early Universe is however limited by several requirements and observations that need to be checked carefully. We are going to list them here \cite{BGGV22003}.

First of all, the decay of the axion must occur well before the primordial nucleosynthesis or even the baryogenesis. Depending on the specific baryogenesis mechanism, the lower bound may become very strict or somewhat relaxed. Here we adopt the nucleosynthesis bound, which translates into a minimal mass for the axion of 
\begin{equation}
    m>10 TeV \sim 10^{-14} M_P. \label{Constraint nucleosynthesis} 
\end{equation}

Secondly, the axion should not decay before its dominance phase. This means that we need $H_d<H_\sigma$. This constraint is automatically satisfied for the case of the early dominance $\sigma_i >1$. For the late dominance, we have $H_\sigma\sim m \sigma_i^4$ and the constraint requires \begin{equation}
    \sigma_i>\sqrt{\frac{m}{M_P}}. \label{Constraint late dominance}
\end{equation}

On the high-energy side, we do not want the axion to dominate at the beginning of the post-big bang phase: it was a subdominant field in the Pre-Big Bang and thus its energy density should be somewhere below the string scale. Therefore, 
\begin{equation}
    \sigma_i<H_i/m. \label{Constraint early dominance}
\end{equation}

\begin{figure}[t]
 \centering
 \includegraphics[width= \textwidth]{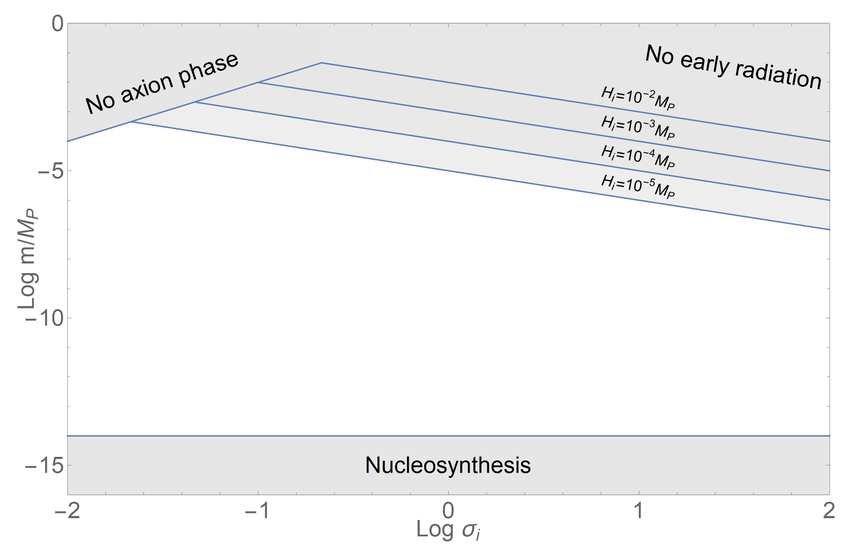}
 \caption{Allowed parameter space in the plane defined by the initial value of the axion in Planck units $\sigma_i$ and its mass $m$. Several lines are drawn depending on the chosen bounce scale $H_i$.}\label{Fig m}
 \end{figure}

Fig. \ref{Fig m} illustrates the space left by these constraints on the background evolution. The requirement of an early radiation phase before the axion dominance depends on the bounce scale $H_i$, but still there is a very wide range of possible masses allowed for the Kalb-Ramond axion and for its initial value \cite{BGGV22003}.

Besides the requirements on the background evolution, we now have very accurate measurements from several cosmological observations: galaxy counts, weak lensing and in particular the CMB anisotropies, which provide the most stringent limits on the amplitude and the slope of the power spectrum. We have already discussed the spectral index of the axion fluctuations, which is transferred to $\zeta$ and $\Psi$ via the curvaton mechanism. 

The amplitude of the axion fluctuations is typically regulated by the ratio of the string scale (coinciding with $H_i$ in our Post-Big Bang history) to the Planck mass. The adiabatic spectrum inherited by the curvature perturbations can be further amplified if the phase preceding the axion dominance is very long. In fact, from Eq. (\ref{Eqzeta'}) we see that any non-adiabatic pressure leads to a growth of $\zeta \sim a^2 \eta^2 \delta p_{nad}$. In the early radiation phase, this corresponds to $\zeta \sim \frac{a_\sigma^4}{a_i} \sigma \chi \sim \frac{\chi}{\sigma_i}$. Therefore, the lower the initial axion value, the higher the amplification taking place before the axion dominates. However, for large values of the axion, the axion-driven inflation amplifies the fluctuations proportionally to $\sigma_i^2$. The interplay between these two effects determines the final amplitude of the adiabatic scalar perturbations encoded in the Bardeen potential as \cite{BGGV22003}
\begin{equation}
    P_\Psi(k) \simeq f^2(\sigma_i)\frac{H_i^2}{M_P^2} \left( \frac{k}{H_i} \right) ^{n_\sigma-1}
\end{equation}
where
\begin{equation}
    f(\sigma_i)=c_1 \sigma_i +\frac{c_2}{\sigma_i} - c_3
\end{equation}
and the constants $c_1=0.13$, $c_2=0.25$, $c_3=0.01$ can be estimated by fitting the spectra obtained by careful numerical simulations.

Comparing the amplitude of this spectrum with the current PLANCK limits \cite{PLANCK2020} of
\begin{equation}
    \mathcal{A}_S=(2.10 \pm 0.03) \times 10^{-9},
\end{equation}
we can put interesting constraints on the bounce scale.

\begin{figure}[t]
 \centering
 \includegraphics[width= \textwidth]{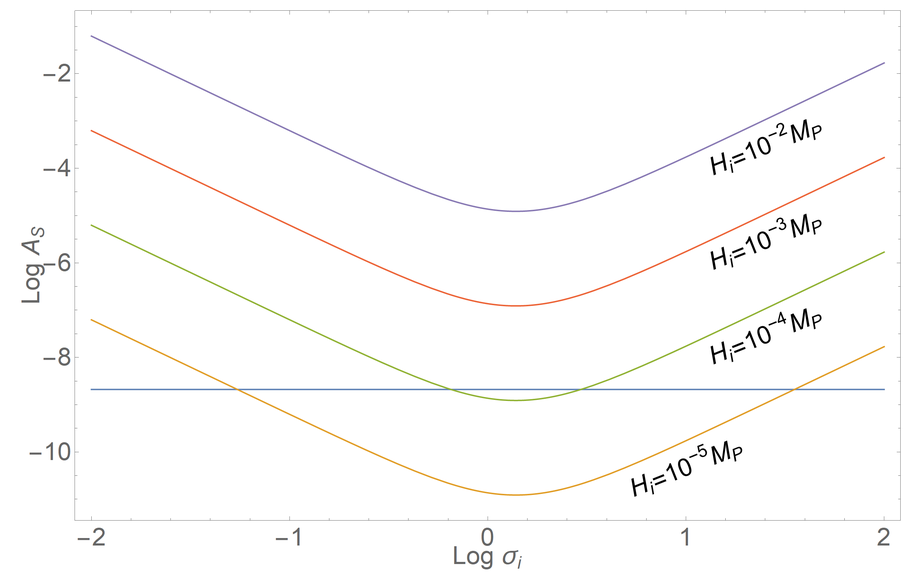}
 \caption{Amplitude of the spectrum of scalar perturbations predicted by the Pre-Big Bang scenario for different values of the bounce scale $H_i$ and the initial axion value in Planck units $\sigma_i$.}\label{Fig AS}
 \end{figure}

Fig. \ref{Fig AS} shows that a bounce scale $H_i=10^{-2} M_P$ generates too large fluctuations, incompatible with the observations. This value is what is naturally expected for the string mass and would be the natural choice for a bounce driven by string effects modifying General Relativity at these scales. However, starting from $H_i=10^{-3.89}$, an amplitude of scalar fluctuations compatible with the PLANCK observations becomes possible.

The mechanism driving the bounce may well start being effective at scales slightly below the string mass. Indeed we do not know enough of the string physics to exclude or validate the possibility that the bounce occurs at scales of the order of $10^{-4} M_P$. Note that this scale is quite close to the GUT scale. Indeed, the observed amplitude of cosmological perturbation indicates that the mechanism for their generation should have something to do with this scale, either through a standard inflation or through a cosmic bounce.

\section{Conclusions}

Standard inflation provides a very simple and predictive model for the early Universe. It solves the curvature and horizon problems and naturally provides a nearly scale-invariant spectrum compatible with the observations. There is currently no compelling need to replace inflation with other models. However, everybody knows that inflation cannot be the end of the story. In particular, what happens to our space-time if we go past beyond the inflationary era remains a theoretical puzzle. Indeed inflation makes the Universe strongly independent of what existed before the accelerated expansion and any possible weirdness in space-time. However, we cannot exclude that any signatures of a quantum Universe survived through inflation and is observable today. Therefore, the investigation of the initial singularity problem and the possible alternatives offered by string theory and other quantum gravity alternatives remain important now as ever. Furthermore, although the current cosmological picture is overall very robust, some minor inconsistencies and tensions might actually be serious signals of new physics that needs to be addressed. It is possible that some of these details hide the sought signatures of an ancient pre-inflationary era.

Bouncing cosmologies start from the assumption that the Big Bang singularity may just be the outcome of a hazardous extrapolation of General Relativity beyond its domain of validity. A complete theory should avoid the singularity and allow to extend our past history back to a pre-bounce growing curvature phase. Among the possible bouncing cosmologies, here we have focused on the Pre-Big Bang scenario, which uses the fields available in bosonic sector of the heterotic string theory to build a complete history of the Universe, from an asymptotically flat past to the current expansion phase. 

The main issue of bouncing cosmologies is that they generally predict steep blue power spectra for scalar perturbations, in contrast with all observations of the Cosmic Microwave Background, galaxy counts and large-scale structures. However, we have seen at least three ways to obtain a scale-invariant spectrum. 
\begin{itemize}
    \item A dust-dominated contraction would generate a scale-invariant spectrum in the pre-bounce era, but it is possibly exposed to background instabilities.
    \item A slow contraction would generate a scale-invariant spectrum in the growing mode, which is matched to a decaying mode in the Post-Big Bang unless some unknown physics intervenes at the bounce and allows the transfer of the spectrum to the constant mode. 
    \item The curvaton mechanism may convert the initially isocurvature fluctuations of some subdominant fields to adiabatic. In the case of the Pre-Big Bang scenario, the Kalb-Ramond axion develops a scale-invariant spectrum for a particularly symmetric Pre-Big Bang evolution and is able to pass all observational and theoretical constraints, provided the bounce scale is at $10^{-4}M_P$ or below.
\end{itemize} 

In addition to these, we should also mention cosmological models in which perturbations are not the outcome of the amplification of vacuum fluctuations, but have a thermal origin such as the string gas cosmology \cite{BrandVafa1989,Nayeri2006}. The detailed realization of this scenario, however, is made difficult by our ignorance of string theory. 

All these proposals indicate more convoluted routes to explain what standard inflation predicts without apparent difficulties. However, we already know that inflation is too simple to be the end of the story and that something is hidden behind it. And even more intriguing, the existence of some yet unsolved tensions warns us that the Universe is indeed more complicated than the standard concordance model we hoped for just a few years ago. So, it is not impossible that we need to complicate standard inflation at a similar level as bouncing cosmologies in order to account for all the details.

We may wonder whether any distinctive features exist that may rule out one scenario or the other. Indeed, scalar perturbations are sensitive to contributions from all possible fields contained in the Universe. It is not impossible to obtain an acceptable spectrum in both scenarios. On the other hand, tensor perturbations are dramatically different: a nearly scale-invariant spectrum for inflation contrasts with the steep blue spectrum obtained in all bouncing cosmologies. In this case, it is difficult to imagine any mechanisms that may bring the tensor spectrum back to scale-invariant. Therefore, the discovery of tensor modes in CMB may definitely rule out bouncing scenarios or relegate them to a theoretical UV completion to standard inflation with no observational consequences. Conversely, a missing detection of tensor modes by more and more precise probes may put inflation in a difficult position. Finally, the continual increase in sensitivity of gravitational wave interferometers may lead to the discovery of the blue end of the spectrum of stochastic gravitational waves and provide a surprising validation of bouncing cosmologies.



\acknowledgments{I wish to thank the editors Giovanni Marozzi and Luigi Tedesco for inviting me to contribute to this special issue. I express all my gratitude to Maurizio Gasperini for his gentle and wise guidance in the years of my early career. By supervising my PhD thesis, he introduced me to String Cosmology, String Theory and the exciting world of theoretical physics at the Theory division at CERN, where I then met Gabriele Veneziano and Massimo Giovannini, with whom I developed most of the studies presented in this review. Of course, I always warmly thank my first supervisor Gaetano Scarpetta, who introduced me to Maurizio Gasperini and followed all my steps through physics and astrophysics.}



\reftitle{References}
\bibliography{refs.bib}



\end{document}